\documentclass[aps,prd,preprint,superscriptaddress]{revtex4}
\usepackage{graphicx}
\usepackage{amsmath}
\usepackage{bm}
\usepackage{here}

\begin{document}
\newcommand{\vecvar}[1]{\mbox{\boldmath$#1$}}

\preprint{OU-HET-625}

\title{Transverse momentum distributions of quarks in the
nucleon from the Chiral Quark Soliton Model}


\author{M.~Wakamatsu}
\affiliation{Department of Physics, Faculty of Science, \\
Osaka University, \\
Toyonaka, Osaka 560-0043, JAPAN}



\begin{abstract}
We report the first calculation of the simplest but most fundamental
transverse momentum dependent (TMD) distribution of quarks in the
nucleon, i.e. the time-reversal-even unpolarized TMD quark and
antiquark distribution with isoscalar combination, within the framework
of the chiral quark soliton model. The nonperturbative account of
the deformed Dirac-sea quarks within the theoretical scheme enables
us to make a reliable predictions not
only for the quark distribution but also for the antiquark distribution.
We found that the predicted average transverse momentum square
$\langle k_\perp^2 \rangle$ of quarks and antiquarks depends strongly
on their longitudinal momentum fraction $x$, which means that
the frequently used assumption of factorization in $x$ and $k_\perp$ is
significantly violated. It is also found, somewhat unexpectedly, that
the average transverse momentum square of antiquarks is considerably
larger than that of quarks.

\end{abstract}


\maketitle


\section{Introduction}

Growing attention has recently been paid to the
transverse-momentum-dependent
(TMD) parton distributions also called the unintegrated parton
distributions \cite{CS1982}.
The reason of particular interest stems from the
fact that they play important roles in the theoretical
description of single spin asymmetries in various hard processes
including semi-inclusive deep inelastic scatterings, Drell-Yan
processes, etc. \cite{SMC2000}\nocite{HERMES2005}-\cite{COMPASS2005}.
The TMD parton distributions are interesting themselves also
from more general viewpoint. They are expected
to give a three-dimensional view of the parton distributions
in momentum space, thereby providing us with
complementary information besides what can be obtained
through generalized parton distributions \cite{Burkardt2000}
\nocite{RP2002}\nocite{Diehl2002}-\cite{BJY2004}.

It has been established that at the leading twist there are
totally eight TMD parton distributions \cite{MT1996}
\nocite{GMS2005}-\cite{BDGMMS2007}. Among those,
six are even under naive time-reversal transformations
(for brevity, we call it T-even), while the remaining two
are T-odd \cite{Sivers1990}\nocite{Sivers1991}-\cite{BM1998}.
To meet the general requirement of factorization and
gauge-invariance, these TMD parton distributions must be
defined with the gauge link operators called Wilson
lines \cite{BHS2002}\nocite{Collins2002}\nocite{Collins1998}
\nocite{JY2002}\nocite{BJY2004}-\cite{Collins2003}.
Since these Wilson lines, which can be
interpreted as simulating the effects of gluon initial and
final state interactions, are process
dependent \cite{Collins2002}, the TMD parton distributions
are in general non-universal \cite{BMP2006},\cite{BM2008}.
Still, a remarkable difference between the T-even and
T-odd TMD parton distributions should not be overlooked.
For the former, the presence of the Wilson lines is less
crucial in the sense that they {\it do} survive even without
such gauge links. On the contrary, for the T-odd TMD
parton distributions, the introduction of the Wilson lines
is of fatal importance since they {\it do} vanish
identically in the absence of such effects of initial and
final state interactions \cite{BHS2002},\cite{Collins2002}. 
The existence of the former types of TMD distributions
appears only natural since the intrinsic motion of quarks
described by the naive bound state wave
function without any complex phase is likely to be
three dimensional and it should be described by some
TMD distribution function. 

Unlike the integrated parton distributions,
our knowledge on the TMD parton distributions is very poor.
Very recently, through the analysis of azimuthal asymmetries
both in hadron-hadron collisions and in semi-inclusive
deep-inelastic scatterings, Anselmino et al. extracted
a T-odd TMD distribution $f_{1 T}^\perp$ called the Sivers
function \cite{ABCDEGKMMMPSVY2005}.
There also exists an attempt to extract another
T-odd distribution called the Boer-Mulders
function \cite{ZLMS2008}.
In these studies, however, a flavor independent
Gaussian distribution of the transverse momentum under the
factorized ansatz in the variables $x$ and $k_\perp$ is
assumed by reason of simplicity, although there is no
compelling reason for this choice.
Since enough empirical information to test such assumption
cannot be expected at the present moment, any information
from models of baryons and/or the lattice QCD would be
extremely valuable.

Although there exist a lot of model calculations for the
integrated parton distributions, there are not so many for the
TMD distributions. (See \cite{BCR2008} for a compact overview
of those studies.) The leading-twist T-even functions were
calculated in a spectator model with scalar and axial-vector
quarks \cite{JMR1997}, and in a light-cone quark model where
the Fock expansion is truncated to consider only 3 valence
quarks \cite{PCB2008}.
On the other hand, the T-odd functions are investigated in the
spectator model with scalar diquarks \cite{GGO2003},\cite{BBH2003}
and with scalar and axial-vector diquarks \cite{GGS2008},
\cite{BSY2004},\cite{BCR2008} in the MIT bag model \cite{Yuan2003}
\nocite{CAKM2006}-\cite{CSV2008},
and in the constituent quark model \cite{CFSV2008}.
Very recently, the first lattice QCD
calculation of the lowest moments of the TMD quark distributions
$f_1 (x,\bm{k}_\perp)$ and $g_{1 T} (x,\bm{k}_\perp)$ has also
been reported \cite{LHPC2008}. (Although our main interest here
is concerned with the TMD quark distributions inside the nucleon,
we recall that there also exist some investigations
on the TMD distributions in the pseudoscalar mesons including the
pion based on the SU(3) Nambu-Jona-Lasinio model \cite{DA1995}
\nocite{WAG1999}-\cite{Arriola2002}.)

Now, the purpose of the present paper is to evaluate the simplest
but most fundamental TMD parton distributions 
in the nucleon, i.e. the T-even unpolarized TMD quark and
antiquark distributions with isoscalar combination, within
the framework of the chiral quark soliton 
model (CQSM) \cite{DPP1988},\cite{WY1991}.
The greatest advantage of the CQSM over the other models of
baryons is its full account of nonperturbative chiral
dynamics of QCD, which has been proven to be essential
for the physics of light
quark distribution functions \cite{DPPPW1996}\nocite{DPPPW1997}
\nocite{WGR1997A}\nocite{WGR1997B}\nocite{WK1998}
\nocite{WK1999}-\cite{Wakamatsu2003}.
The nonperturbative account of the
deformed Dirac-sea quarks within the scheme enables us to make
a reliable predictions not only for the quark distributions but
also for the antiquark distributions, as already been proven
through its good reproduction of the famous
NMC (New Muon Collaboration) observation, i.e.
the dominance of the $\bar{d}$ sea over the $\bar{u}$ one in the
proton, etc. \cite{Wakamatsu1992},\cite{WK1998},\cite{PPGWW1999}.
We therefore expect that the same model would
provide us with valuable and reliable information also on the nature
of the transverse motion of quarks and antiquarks inside the nucleon.

The plan of the paper is as follows. First, in sect.II, we
derive theoretical formulas necessary for the calculation of
the TMD quark and antiquark distributions within the framework
of the CQSM. Next, in sect.III, the predictions of the CQSM are
shown and the detailed discussion on them will be given.
Conclusion of our analysis is then given in sect.IV.

\section{Unpolarized TMD quark distributions \label{Sect:formalism}}

We are interested here in the simplest but most fundamental
TMD parton distribution function, i.e. the unpolarized
TMD quark distribution. The unpolarized TMD quark distribution
function of flavor $a$ in the nucleon, averaged over its
spin, is defined as
\begin{equation}
 f^a (x, \bm{k}_\perp) \ = \ 
 \int \frac{d \xi^- \,d^2 \bm{\xi}_\perp}{2 \,(2 \,\pi)^3} \,\,
 e^{i \,k \cdot \xi} \,\langle P \,| \,\left.
 \bar{\psi}_a (0) \,(1 + \gamma^0 \,\gamma^3) \,
 {\cal U}_{[0,\xi]} \,\psi_a (\xi) \,| \,P \rangle \,
 \right|_{\xi^+ = 0} , \label{Eq:TMDpdf}
\end{equation}
where
\begin{equation}
 {\cal U}_{[0,\xi]} \ \equiv \ {\cal P} \,
 e^{- \,i \,g \,\int_0^\xi \,A (w) \cdot d w} ,
\end{equation}
is the so-called gauge link operator, also called Wilson line,
connecting the two different space-time points $0$ and $\xi$,
by all possible ordered path.
This gauge link is known to simulate the interaction of the
outgoing quark field with the spectators inside the hadron,
and it works to ensure the color gauge invariance of the
above definition of the TMD quark distribution \cite{Collins2002}.

Here, we want to evaluate the TMD quark distribution within
the framework of the CQSM. Since the CQSM is an effective
quark theory, which does not contain gluonic degrees of
freedom at least explicitly, we can drop the gauge link
operator. As a matter of course, this simple procedure is
fatal for predicting the T-odd TMD distributions like the Sivers
function \cite{Sivers1990},\cite{Sivers1991} or the Boer-Mulders
function \cite{BM1998}, since the complex
phase arising from the gauge link is indispensable for
the existence of such T-odd distribution functions.
Roughly speaking, the T-even distribution function we are
to calculate below can be thought of as a ``static''
TMD quark distribution, which can be computed directly from the
bound state wave functions of a target nucleon without accompanying
complex phase arising from the effects of final state
interactions \cite{Brodsky2009}.

Though the definition (\ref{Eq:TMDpdf}) of the TMD distribution
function itself is Lorentz-frame independent, it is convenient
to evaluate it in the nucleon rest frame, where $\bm{P} = 0$
and $P_0 = M_N$ with $M_N$ being the nucleon mass.
This enables us to rewrite the exponential factor
$e^{\,i \,k \cdot \xi}$ contained in (\ref{Eq:TMDpdf})
in the following manner :
\begin{equation}
 e^{\,i \,k \cdot \xi} \ = \ 
 e^{\,i \,k_+ \,\xi^- \,- \,i \,\bm{k}_\perp \cdot \bm{\xi}_\perp}
 \ = \ 
 e^{\,i \,x \,P_+ \,\xi^- \,- \,i \,\bm{k}_\perp \cdot \bm{\xi}_\perp}
 \ = \
 e^{\,i \,x \,M_N \,\xi^0 \ - \ i \,\bm{k}_\perp \cdot \bm{\xi}_\perp}.
\end{equation}
Here, we have used the relation $\xi^3 = - \,\xi^0$
(or $\xi^+ = 0$), together with the standard definition of the
light-cone vector, $\xi^\pm = (\xi^0 \pm \xi^3) \,/ \,\sqrt{2}$.

To proceed further, we notice that the large-$N_c$ behavior of
the unpolarized TMD quark distribution depends on the
isospin combination \cite{Pobylitsa2003}.
Just similarly to the integrated
unpolarized distributions extensively studied
before \cite{DPPPW1996}\nocite{DPPPW1997}-\cite{WGR1997A},
\cite{WK1998},\cite{Wakamatsu2003},
the isoscalar distribution
is dominant in the large-$N_c$ limit, and survives
at the mean-field level, while
the isovector combination appears as the first order
correction in the collective angular velocity operator
$\Omega$ of the rotating soliton.
(We recall that $\Omega$
is an $1 / N_c$ quantity.) Since the evaluation of the
$O (\Omega^1)$ contribution is much harder,
let us concentrate on the calculation of the isoscalar
part in the present paper. Using the formalism developed in
the previous studies \cite{DPPPW1996}\nocite{DPPPW1997}
\nocite{WGR1997A}\nocite{WGR1997B}\nocite{WK1998}
\nocite{WK1999}-\cite{Wakamatsu2003},
we find that the isoscalar unpolarized
TMD quark distribution $f^{u+d} (x,\bm{k}_\perp)$ in the CQSM
is given in the following form : 
\begin{eqnarray}
 f^{u+d} (x,\bm{k}_\perp) &=& N_c \,M_N \,
 \int \,d^3 \bm{R} \,\int \,
 \frac{d \xi^0 \,d^2 \bm{\xi}_\perp}{(2 \,\pi)^3} \,\,
 e^{\,i \,x \,M_N \,\xi^0 \,- \,i \,\bm{k}_\perp \cdot \xi_\perp}
 \nonumber \\
 &\,& \hspace{6mm} \times \left.
 \sum_{n \in occ} \,e^{\,i \,E_n \,\xi^0} \,
 \Phi^\dagger_n (- \bm{R}) \,( 1 + \alpha^3 ) \,
 \Phi_n (\bm{\xi} - \bm{R}) \,\right|_{\xi^3 = - \xi^0} ,
 \label{Eq:TMDpdf_CQSM_1}
\end{eqnarray}
with $\alpha^3 = \gamma^0 \,\gamma^3$. 
Here, $\Phi_n (\bm{x})$ is the eigenfunctions of the Dirac
hamiltonian $H$ with the corresponding eigenvalue $E_n$, i.e. 
\begin{equation}
 H \,\Phi_n (\bm{x}) \ = \ E_n \,\Phi_n (\bm{x}) ,
\end{equation}
with
\begin{equation}
 H \ = \ \frac{\bm{\alpha} \cdot \nabla}{i} \ + \ 
 M \,\beta \,e^{\,i \,\gamma_5 \,\bm{\tau} \cdot \hat{\bm{r}} \,F(r)}.
\end{equation}
In (\ref{Eq:TMDpdf_CQSM_1}), $sum_{n \in occ}$ denotes the summation
over the occupied states in the hedgehog mean field.
In deriving (\ref{Eq:TMDpdf_CQSM_1}), the projection into a nucleon
state with given momenta
$\bm{P}$ has been achieved, as usual, by integrating over all shift of
the center-of-mass coordinate $\bm{R}$ of the soliton,
\begin{equation}
 \langle \bm{P}^\prime \,| \,\cdots \,| \,\bm{P} \rangle \ = \ 
 \int \,d^3 \bm{R} \,\,e^{\,i \,(\bm{P}^\prime - \bm{P}) \cdot \bm{R}}
 \,\,\cdots .
\end{equation}
Introducing the eigenfunctions in the momentum representation,
\begin{equation}
 \Phi_n (\bm{x}) \ = \ \int \,\frac{d^3 \bm{p}}{(2 \,\pi)^3} \,\,
 e^{\,i \,\bm{p} \cdot \bm{x}} \,\,\tilde{\Phi}_n (\bm{p}),
\end{equation}
(\ref{Eq:TMDpdf_CQSM_1}) can readily be transformed into the form :
\begin{eqnarray}
 f^{u+d} (x,\bm{k}_\perp) &=& N_c \,M_N \,\int \,
 \frac{d \xi^0 \,d^2 \bm{\xi}_\perp}{(2 \,\pi)^3} \,
 e^{\,i \,x \,M_N \,\xi^0 \,- \,i \,\bm{k}_\perp \cdot \bm{\xi}_\perp}
 \nonumber \\
 &\,& \times \ \sum_{n \in occ} \,e^{i \,E_n \,\xi^0} \,\,
 \int \,\frac{d^3 \bm{p}}{(2 \,\pi)^3} \,\,
 e^{- \,i \,\bm{p} \cdot \bm{\xi}} \,\,
 \tilde{\Phi}_n^\dagger (\bm{p}) \,(1 + \alpha_3) \,
 \tilde{\Phi}_n (\bm{p}) .
\end{eqnarray}
Now, by noting that 
$e^{\,i \,\bm{p} \cdot \bm{\xi}} \ = \ 
e^{\,i \,p^3 \,\xi^3} \,e^{\,i \,\bm{p}_\perp \cdot \bm{\xi}_\perp} \ = \ 
e^{\,- \,i \,p^3 \,\xi^0} \,e^{\,i \,\bm{p}_\perp \cdot \bm{\xi}_\perp}$,
one can carry out the integration
over $\xi^0$ and $\bm{\xi}_\perp$ to obtain
\begin{eqnarray}
 f^{u+d} (x,\bm{k}_\perp) &=& M_N \,N_c \,
 \int \,\frac{d^3 \bm{p}}{(2 \,\pi)^3} \nonumber \\
 &\,& \times \ \sum_{n \in occ} \,
 \tilde{\Phi}_n^\dagger (\bm{p}) \,(1 + \alpha^3) \,
 \delta ( x \,M_N - E_n - p^3) \,\delta^2 (\bm{p}_\perp - \bm{k}_\perp) \,
 \tilde{\Phi}_n (\bm{p}) . \label{Eq:TMDpdf_CQSM_2}
\end{eqnarray}
As usual, the numerical evaluation of the above expression is
carried out by using the discretized momentum basis of Kahana and
Ripka \cite{KR1984},\cite{KRS1984},
so that it is convenient to introduce the smeared distribution
defined by \cite{DPPPW1997}
\begin{equation}
 f^{u+d}_\gamma (x, \bm{k}_\perp) \ \equiv \ 
 \frac{1}{\sqrt{\pi} \,\gamma} \,\int_{- \,\infty}^\infty
 \,d x^\prime \,\,
 e^{- \,\frac{(x - x^\prime)^2}{\gamma^2}} \,\,
 f^{u+d} (x,\bm{k}_\perp) . \label{Eq:TMDpdf_smear}
\end{equation}
After some straightforward algebra, we therefore get
\begin{eqnarray}
 f^{u+d}_\gamma (x, \bm{k}_\perp) &=& N_c \,\sum_{n \in occ} \,
 \int_{- \,\infty}^\infty \,d k_3 \,\,
 e^{- \,\frac{1}{\gamma^2} \,
 \left( x - \frac{E_n + k_3}{M_N} \right)^2} \,
 \tilde{\Phi}_n^\dagger (\bm{k}_\perp, k_3) \,(1 + \alpha^3) \,
 \tilde{\Phi}_n (\bm{k}_\perp, k_3) .
\end{eqnarray}
Here, we have changed the integration variable from $p^3$ to $k_3$.
We recall now that $\tilde{\Phi}_n$ is simultaneous eigenfunctions
of the hamiltonian $H$, and the grand spin operator
$\bm{K} = \bm{J} + \frac{1}{2} \,\bm{\tau}$ and its projection
$M_K$ on the $z$-axis. The functions $\tilde{\Phi}_n$ can be
expanded in terms of the discretized plane-wave basis of Kahana
and Ripka as
\begin{equation}
 \tilde{\Phi}_n \ = \ \sum_\alpha \,\sum_i \,c^{(n)}_{\alpha i} \,
 \phi_{\alpha i} ,
\end{equation}
where the index $\alpha$ distinguishes 4 independent plane-wave basis
with definite grand spin $K, M_K$, and parity. (We point out that
the expansion coefficients $c^{(n)}_{\alpha i}$ can be chosen
real numbers.)
On the other hand, $i$ labels the discretized momenta $k_i$,
which is determined by imposing the boundary condition
\begin{equation}
 j_K (k_i \,D) \ = \ 0,
\end{equation}
at the radius $r = D$ chosen to be sufficiently larger than the
typical soliton size. (The number of momentum bases is made finite
by introducing the maximum momentum $k_{max}$ such that
$k_i < k_{max}$.) 
Since the Kahana-Ripka plane-wave basis is given in the spherical
representation, we also express $\bm{k}_\perp$ and $k_3$ in the
spherical coordinates as
\begin{eqnarray}
 \left( \bm{k}_\perp \right)_x &=& k_i \,\sin \theta \,\cos \phi, \\
 \label{Eq:spherical_x}
 \left( \bm{k}_\perp \right)_y &=& k_i \,\sin \theta \,\sin \phi, \\
 \label{Eq:spherical_y}
 k_3 &=& k_i \,\cos \theta . \label{Eq:spherical_z}
\end{eqnarray}
Since the final answer for the TMD distribution $f^{u+d} (x,\bm{k}_\perp)$
should be independent of the azimuthal angle $\phi$ of the
transverse momentum $\bm{k}_\perp$,
the following replacement is justified
\begin{equation}
 \phi_{\alpha i}^\dagger \,(1 + \alpha_3) \,\phi_{\beta j}
 \ \longrightarrow \ \int \,\frac{d \phi}{2 \,\pi} \,
 \phi_{\alpha i}^\dagger \,(1 + \alpha_3) \,\phi_{\beta j} .
\end{equation}
The result of $\phi$ integration can schematically be written as
\begin{equation}
 \int \,\frac{d \phi}{2 \,\pi} \,\,
 \phi_{\alpha i}^\dagger \,(1 + \alpha_3) \,\phi_{\beta j} \ = \ 
 \delta_{i \,j} \,\,{\cal F}_{\alpha \,\beta} (k_i, \cos \theta),
\end{equation}
where ${\cal F}_{\alpha \,\beta} (k_i, \cos \theta)$ is a function
of $k_i$ and $\cos \theta$.
The Kronecker delta $\delta_{i j}$ arises from the
fact that the operator $(1 + \alpha_3)$ does not change momentum.
Because the magnitude of the transverse momentum
$k_\perp \equiv | \bm{k}_\perp |$
is given externally, the relations 
(\ref{Eq:spherical_x})-(\ref{Eq:spherical_z}) dictates that, once the plane-wave-basis momentum $k_i$ is given, both of
$k_3$ and $\cos \theta$ are fixed as
\begin{eqnarray}
 k_3 &=& \pm \,\sqrt{k_i^2 - k_\perp^2} \ \equiv \ k_{3,i}, \ \ \ \ \ 
 \cos \theta \ = \ \frac{k_3}{k_i} \ \equiv \ \cos \theta_i.
\end{eqnarray}
We therefore approximate the expression (\ref{Eq:TMDpdf_smear}) as
\begin{eqnarray}
 f^{u+d}_\gamma (x,\bm{k}_\perp) \ = \ N_c \,\sum_{n \in occ} \,
 \sum_{\alpha, \beta} \,\sum_i \,
 c^{(n)}_{\alpha i} \,c^{(n)}_{\beta i} \,\,
 {\cal F}_{\alpha \beta} (k_i, \cos \theta_i) \,
 \frac{1}{\sqrt{\pi} \,\gamma}
 e^{- \,\frac{1}{\gamma^2} \,
 \left( x - \frac{E_n + k_{3,i}}{M_N} \right)^2} .
\end{eqnarray}
In the next section, we show that this algorithm in fact works if
we take large enough boundary radius $D$, at least for the calculation
of Dirac sea contributions. Unfortunately, we find that the same
algorithm does not work very well for the calculation of the
discrete valence level contribution (corresponding to $n = 0$).
This causes no serious problem,
however, since we can use simpler method to evaluate this part.
We just go back to the expression (\ref{Eq:TMDpdf_CQSM_2})
before introducing smeared distribution, and obtain
\begin{eqnarray}
 f^{u+d}_{val} (x, \bm{k}_\perp) \ = \ M_N \,N_c \,\,
 \tilde{\Phi}_0^\dagger (\bm{k}_\perp, k_3 = x \,M_N - E_0) \,
 (1 + \alpha_3) \,
 \tilde{\Phi}_0 (\bm{k}_\perp, k_3 = x \,M_N - E_0) ,
\end{eqnarray}
where $\tilde{\Phi}_0$ and $E_0$ are the eigenfunction and the eigenenergy
corresponding the valence level. Since $\tilde{\Phi}_0$
is a discrete bound state wave function anyhow, we can evaluate the
above expression without any difficulty by using its momentum space
wave function as used in \cite{WGR1997B}.

\section{Numerical results and discussions \label{Sect:results}}

The basic lagrangian of the model contains two physical
parameters, the weak pion decay constant $f_\pi$, the
dynamically generated effective quark mass $M$.
As usual, $f_\pi$ is fixed to be its physical value,
i.e. $f_\pi = 93 \,\mbox{MeV}$.
On the other hand, $M$ is taken to be $375 \,\mbox{MeV}$,
which is favored from our previous analysis of
the nucleon spin structure functions \cite{WK1999},\cite{Wakamatsu2003}.

The model contains ultraviolet divergences so that it must be
regularized by introducing some physical cutoff. Following the
previous studies, we simply use the Pauli-Villars regularization
scheme with single subtraction.
In this scheme, any nucleon observables including quark distribution
functions in the nucleon are regularized through the
subtraction \cite{DPPPW1997} :
\begin{equation}
 {\langle O \rangle}^{reg} \ \equiv \ {\langle O \rangle}^M \ - \ 
 {\left( \frac{M}{M_{PV}} \right)}^2 \,{\langle O \rangle}^{M_{PV}} .
 \label{PVreg}
\end{equation}
Here ${\langle O \rangle}^{M}$ denotes the nucleon matrix element of
an operator $O$ evaluated with the original effective action with the
mass parameter $M$, while ${\langle O \rangle}^{M_{PV}}$ stands for the
corresponding matrix element obtained from ${\langle O \rangle}^{M}$
by replacing the parameter $M$ with the Pauli-Villars cutoff mass
$M_{PV}$. 
Demanding that the regularized action reproduces
the correct normalization of pion kinetic term in the corresponding
bosonized action, $M_{PV}$ is uniquely fixed by the relation
\begin{equation}
 \frac{N_c}{4 \,\pi^2} \,M^2 \,\log \,\frac{M_{PV}^2}{M^2} \ = \ 
 f_\pi .
\end{equation}
For $M = 375 \,\mbox{MeV}$, this gives $M_{PV} \simeq 562 \,
\mbox{MeV}$, leaving no adjustable parameter.

A short comment may be necessary for the regularization scheme
explained above.
As was shown in \cite{KWW1999}, the Pauli-Villars scheme with
a single subtraction term is not a completely satisfactory
regularization procedure. It fails to remove ultraviolet divergences
of some special quantities like the vacuum quark condensate, which
contains quadratic divergence instead of logarithmic one.
For obtaining finite answers also for these special observables,
the single-subtraction Pauli-Villars scheme is not enough.
It was shown that more sophisticated Pauli-Villars
scheme with two subtraction terms meets this requirement \cite{KWW1999}.
Fortunately, the self-consistent solution of the CQSM obtained in this
double-subtraction Pauli-Villars scheme is only slightly different
from that of the naive single-subtraction scheme,
except when dealing with some special quantities containing quadratic
divergences \cite{KWW1999}.
Considering the fact that the calculation of
quark distribution functions, much more the TMD quark distribution
functions, in the CQSM, is extremely time-consuming
and that the most nucleon observables are rather insensitive to
which regularization scheme is chosen, we shall simply use here
the single-subtraction Pauli-Villars scheme.
(The use of more time-consuming double subtraction scheme is
mandatory, however, for some special parton distribution
functions containing quadratic divergences. An example is
the chiral-odd twist-3 unpolarized distribution functions $e(x)$
investigated in \cite{ES2003}\nocite{WO2003}
\nocite{OW2004}-\cite{COSU2008}.)

\begin{figure}[H] \centering
\begin{center}
 \includegraphics[width=15.0cm]{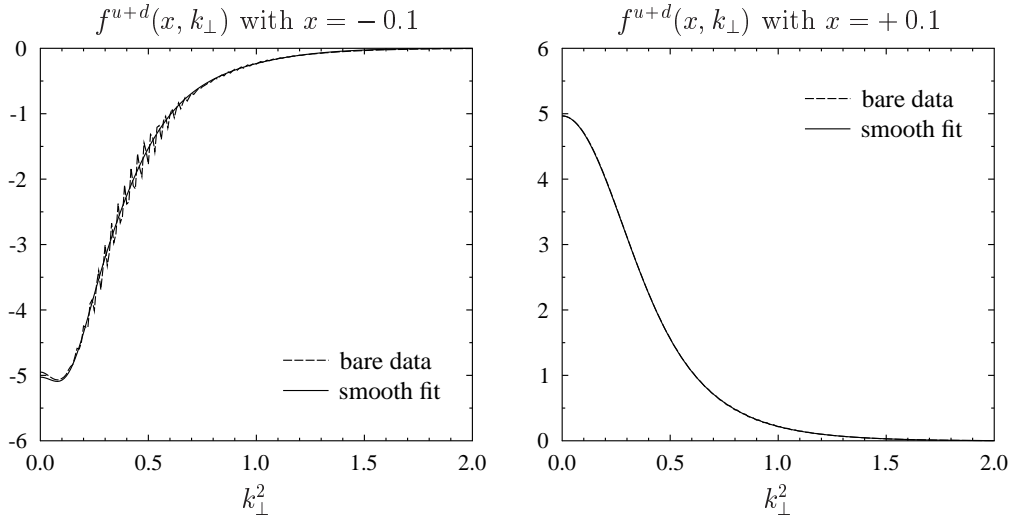}
\end{center}
\vspace*{-1.0cm}
\caption{The Dirac-sea contributions to the unpolarized TMD 
distribution with isoscalar combination $f^{u+d} (x,\bm{k}_\perp)$
for two typical values of $x$, i.e. $x = - \,0.1$ (left panel),
and $x = + \,0.1$ (right panel). In both figures,
the dashed curves represent the bare numerical
predictions of the model obtained with $D = 40 \,/\, M$,
$k_{max} = 10 \,M$, and $\gamma = 0.3$, while the solid curves
are their smooth fits. Note that the dashed and solid curves are
almost indistiguishable in the right panel.}
\label{Fig:bare}
\end{figure}%

First, let us check whether the numerical algorithm proposed in
the previous section in fact works. As pointed out there,
there is no difficulty in the calculation of discrete valence level
contribution to the TMD quark distribution, so that we concentrate
on the Dirac-sea contributions. 
Shown in Fig.\ref{Fig:bare} are the predictions
of the CQSM for the Dirac-sea contributions to
$f^{u+d} (x,\bm{k}_\perp)$ for two typical values of longitudinal
momentum fraction, i.e. $x = - \,0.1$ (left panel) and
$x = + \,0.1$ (right panel).
In these figures, the dashed curves are the bare numerical
predictions of the model obtained with $D = 40 \,/\, M$,
$k_{max} = 10 \,M$, and $\gamma = 0.03$.
The observed fluctuating behavior, which is a little stronger
for the negative value of $x$, is due to the use of the
discretized basis with fairly small $\gamma$.
We fit these results with smooth function shown by
solid curves.

After demonstrating that our numerical method in fact works as
long as one takes large enough boundary radius $D$,
let us now show the full CQSM predictions for the unpolarized
TMD quark and antiquark distribution functions. 
We recall that the function $f^{u+d} (x,\bm{k}_\perp)$ with
positive $x$ can literally be interpreted as quark distribution,
whereas the function with negative $x$ should be interpreted
as antiquark distributions with an extra minus sign as
\begin{equation}
 f^{u+d} (-x,\bm{k}_\perp) \ = \ - \,
 f^{\bar{u}+\bar{d}} (x,\bm{k}_\perp),
 \hspace{10mm} (0 < x < 1) . \label{Eq:negative_x}
\end{equation}

\begin{figure}[H] \centering
\begin{center}
 \includegraphics[height=19.0cm]{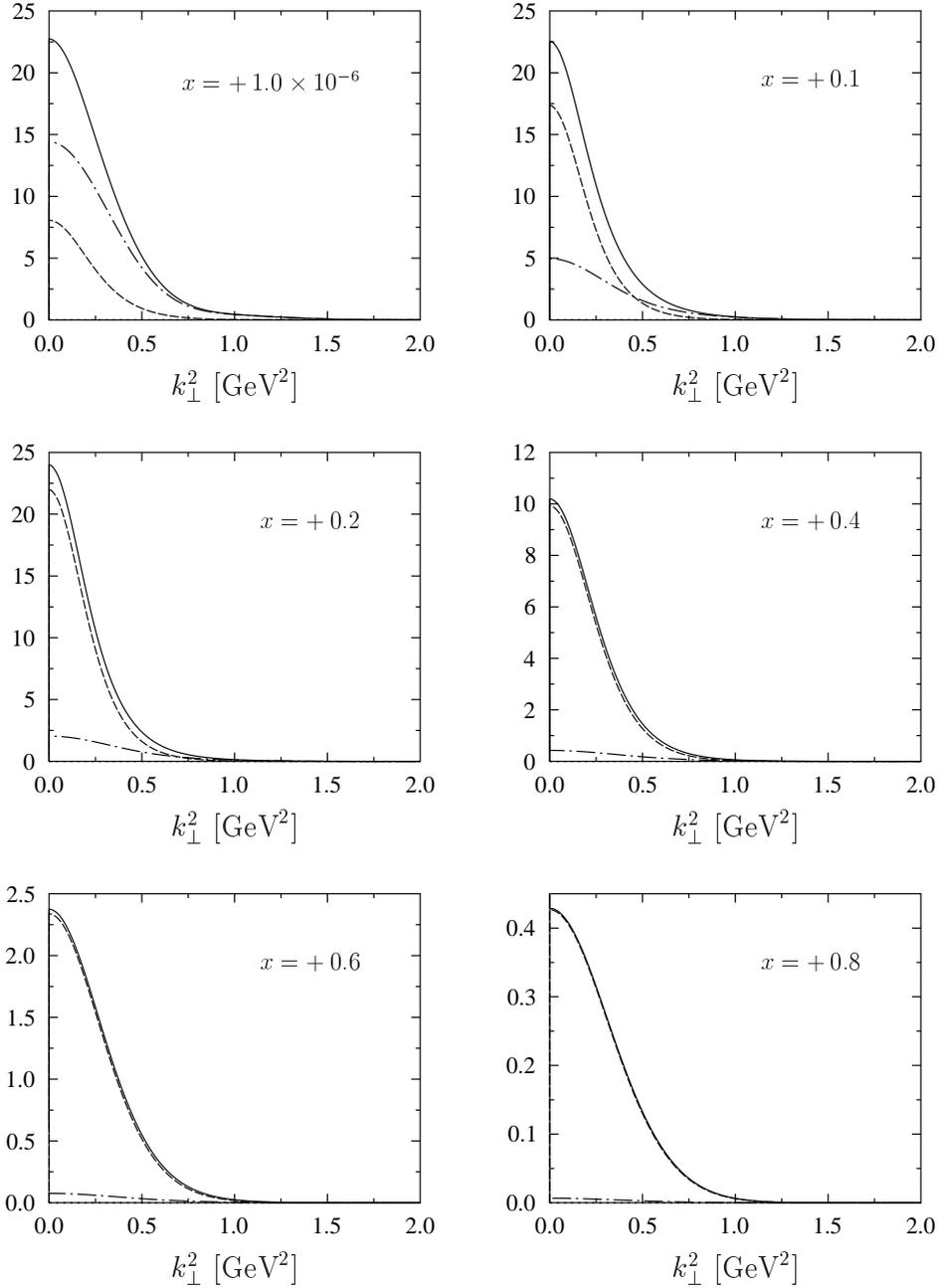}
\end{center}
\vspace*{-1.0cm}
\caption{The CQSM predictions for the unpolarized TMD 
distribution with isoscalar combination $f^{u+d} (x,\bm{k}_\perp)$
in the positive $x$ region. They are shown for 6 typical values
of $x$, i.e. $x = 1.0 \times 10^{-6}, 0.1, 0.2, 0.4, 0.6$ and $0.8$.
In these figures, the dashed and dash-dotted curves respectively
stand for the contributions of the discrete valence level and
of the deformed Dirac-sea quarks, while their sum is represented
by the solid curves.}
\label{Fig:ktf_pos}
\end{figure}

\begin{figure}[H] \centering
\begin{center}
 \includegraphics[height=19.0cm]{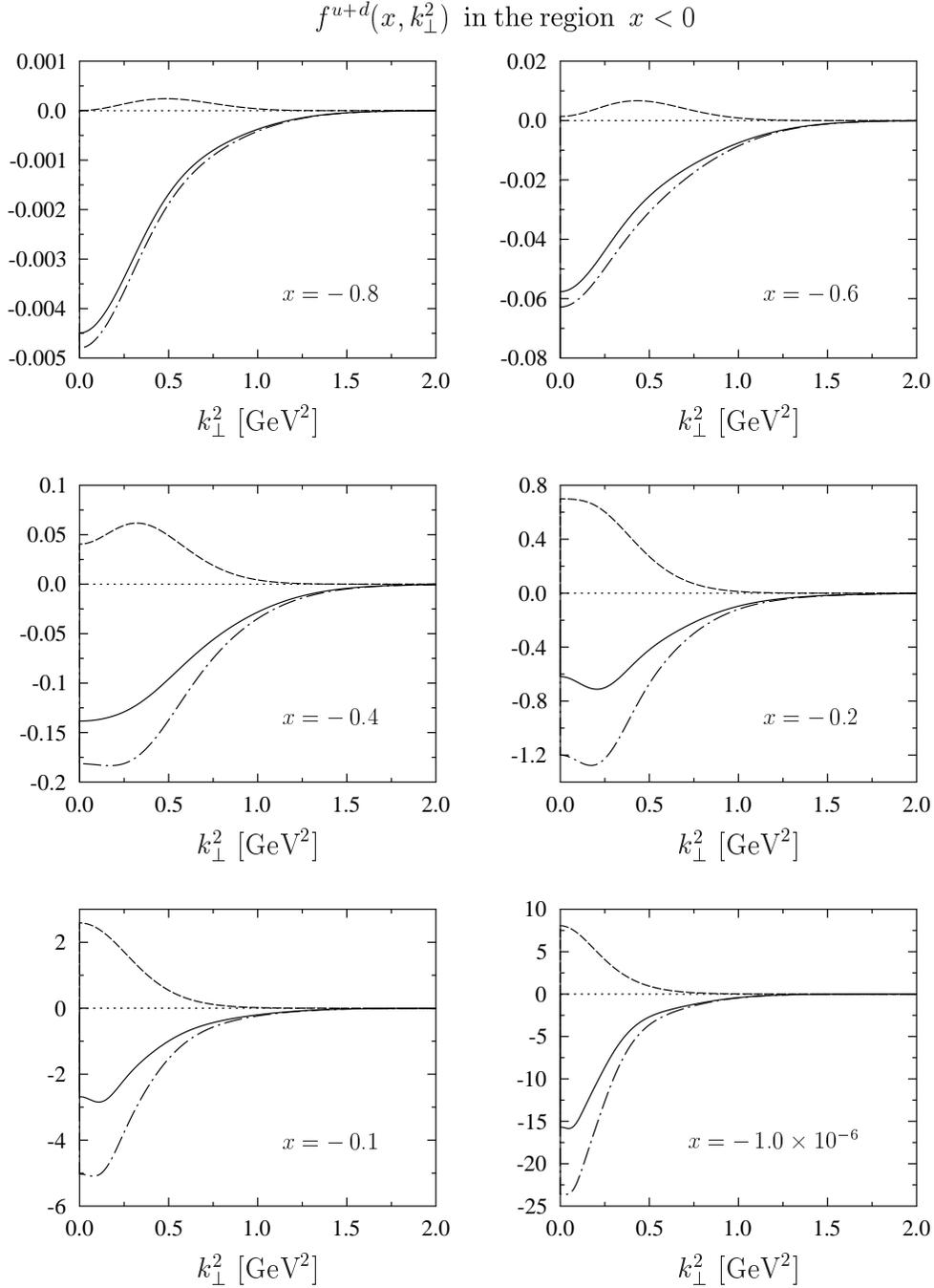}
\end{center}
\vspace*{-1.0cm}
\caption{The CQSM predictions for the unpolarized TMD 
distribution with isoscalar combination $f^{u+d} (x,\bm{k}_\perp)$
in the negative $x$ region, which correspond to the antiquark
distributions, according to the rule
$f^{\bar{u}+\bar{d}} (x, \bm{k}_\perp) = - \,
f^{u+d} (- \,x,\bm{k}_\perp)$ with $0 < x < 1$.
They are shown for 6 typical values
of $x$, i.e. $-0.8, -0.6, -0.4, -0.2, -0.1$
and $- 1.0 \times 10^{-6}$.
The meaning of the curves is the same as in Fig.\ref{Fig:ktf_pos}.}
\label{Fig:ktf_neg}
\end{figure}

Fig.\ref{Fig:ktf_pos} show the CQSM predictions for
$f^{u+d} (x,\bm{k}_\perp)$
as functions of $k_\perp^2$ for 6 typical values of $x$,
i.e. $x = 1.0 \times 10^{-6}, 0.1,
0.2, 0.4, 0.6$ and $0.8$. On the other hand,
Fig.\ref{Fig:ktf_neg} shows
$f^{u+d} (x,\bm{k}_\perp)$ for 6 different values of $x$, i.e.
$x = - \,0.8, - \,0.6, - \,0.4, - \,0.2, - \,0.1$
and $- \,1.0 \times 10^{-6}$.
In these figures, the contributions of the discrete valence-level
quarks and those of the Dirac-sea quarks are illustrated by the
dashed and dash-dotted curves, respectively, while their
sums are denoted by the solid curves.
First, let us look into the quark distribution with $x > 0$.
As expected, the contributions of Dirac-sea quark is most
important in the lower $x$ region, while the contribution from
the discrete valence level dominates over that from the
Dirac-sea quarks as $x$ increases to approach $1$.
Somewhat unexpectedly, the contributions
from the Dirac-sea quarks are seen to have longer range tail
in $k_\perp$ space, as most clearly seen from the figure
corresponding to $x = 0.1$.
(The plot of $f^{u+d} (x,\bm{k}_\perp)$ is not the
best way to see this unique feature.
A better way to see it more clearly
would be to examine the graph of the $k_\perp$-weighted
distribution $k_\perp \,f^{u+d} (x,\bm{k}_\perp)$,
the integral of which over
$k_\perp$ gives the integrated quark distribution
$f^{u+d}(x)$. See the discussion later, for more detail.)

Turning next to the distribution with $x < 0$, corresponding to the
antiquark distribution,
we find that the contributions from the
Dirac-sea quarks dominate over that of valence level.
Here, the fact that the Dirac-sea contributions have higher
$k_\perp$ components than the valence-level quark is much more
evidently seen. An importance notice here is that the total
contributions shown by the solid curves are all negative for any $x$.
In consideration of the extra minus sign indicated in
(\ref{Eq:negative_x}),
this means that the {\it positivity} of the TMD antiquark
distribution $f^{\bar{u} + \bar{d}}
(x,\bm{k}_\perp)$ is legitimately fulfilled in the CQSM.
We emphasize that the proper inclusion of the Dirac-sea contribution
is crucial for the fulfillment of this fundamental property of
unpolarized parton distribution.
It is this unique feature of the CQSM that enables us to make a
reasonable predictions not only for the quark distribution but also
the antiquark (or the sea quark) distributions.

\begin{figure}[htb] \centering
\begin{center}
 \includegraphics[width=16.0cm]{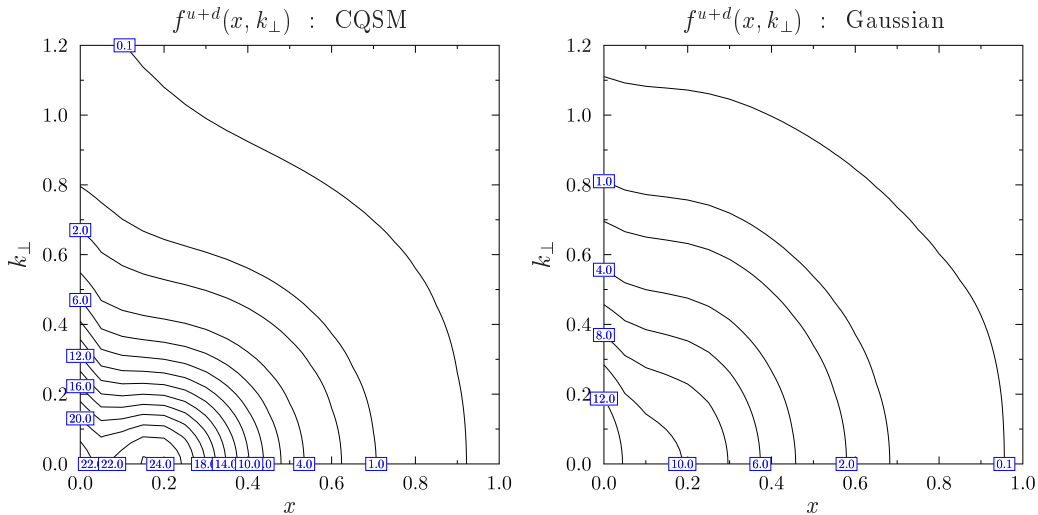}
\end{center}
\vspace*{-1.0cm}
\caption{The contour plot for the isoscalar unpolarized TMD
quark distribution function $f^{u+d} (x,\bm{k}_\perp)$.  
The left panel corresponds to the prediction of the CQSM,
while the right panel to the schematic distribution in the
factorized form $f^{u+d} (x, \bm{k}_\perp) = f^{u+d} (x) \,
\frac{1}{\pi \,\langle k_\perp^2 \rangle} \,
e^{- \,k_\perp^2 \,/ \,\langle k_\perp^2 \rangle}$ with
$\langle k_\perp^2 \rangle = 0.25 \,\mbox{GeV}^2$.}
\label{Fig:fkt2contour_Q}
\end{figure}%

For lack of enough empirical information, the TMD quark
distribution is often assumed in a factorized form with the
Gaussian distribution in $k_\perp$ as
\begin{equation}
 f^{u+d} (x,\bm{k}_\perp) \ = \ f^{u+d} (x) \,\,
 \frac{1}{\pi \,\langle k_\perp^2 \rangle} \,
 e^{\,- k_\perp^2 \,/\, \langle k_\perp^2 \rangle},
 \label{Eq:factorized}
\end{equation}
with $\langle k_\perp^2 \rangle \simeq 0.25 \,\mbox{GeV}^2$.
Then, it may be of some interest to compare the predictions of the
CQSM for the TMD quark distribution $f^{u+d} (x,\bm{k}_\perp)$
with this factorized ansatz. 
We show in Fig.\ref{Fig:fkt2contour_Q} the contour plot
for the isoscalar unpolarized TMD quark distribution function
$f^{u+d} (x,\bm{k}_\perp)$.  
The left panel corresponds to the prediction of the CQSM,
while the right panel to the factorized form given by
(\ref{Eq:factorized}).
(For the integrated distribution
$f^{u+d} (x)$ in the factorized form, we use the prediction
of the CQSM, i.e. it is obtained from the CQSM prediction for
$f^{u+d} (x,\bm{k}_\perp)$ after integration over $k_\perp$.) 
One clearly sees significant difference between the behaviors
of the two distributions especially in the lower $x$ region,
$x < 0.4$, where the magnitude of the TMD distribution is
dominantly large.

\begin{figure}[htb] \centering
\begin{center}
 \includegraphics[width=16.0cm]{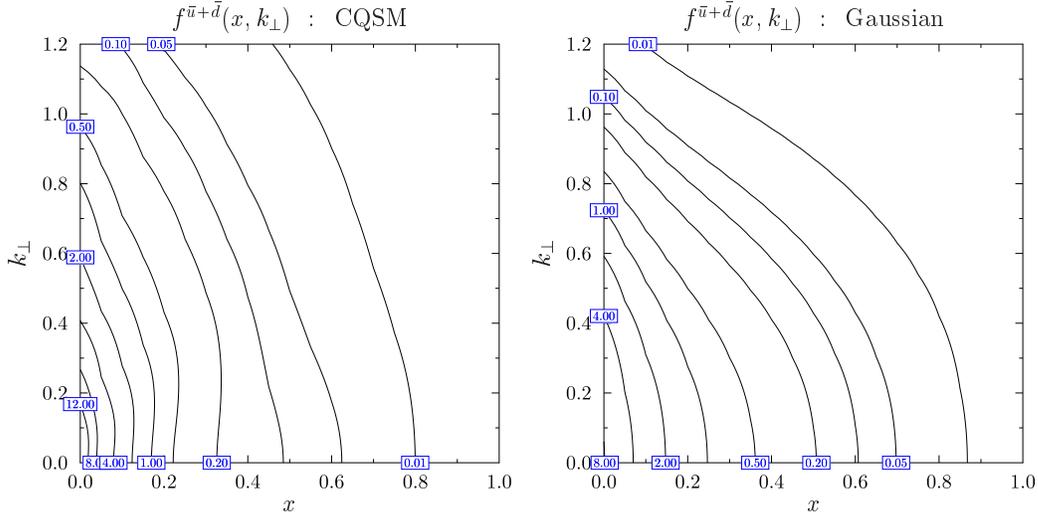}
\end{center}
\vspace*{-1.0cm}
\caption{The contour plot for the isoscalar unpolarized TMD
antiquark distribution $f^{\bar{u}+\bar{d}} (x,k_\perp)$.
The meaning of the curves is the same as
in Fig.\ref{Fig:fkt2contour_Q}.}
\label{Fig:fkt2contour_AQ}
\end{figure}%

Fig.\ref{Fig:fkt2contour_AQ} show a similar comparison
for antiquark distribution
$f^{\bar{u}+\bar{d}} (x,\bm{k}_\perp)$.
One again observes significant difference between the prediction
of the CQSM and the simple parametrization (\ref{Eq:factorized}).
According to the prediction of the CQSM,
the TMD antiquark distribution with lower $x$ is seen to extend
over higher $k_\perp$ region, as compared with the schematic form.
At any rate, the analysis above indicates that the frequently used
assumption of factorization in $x$ and $k_\perp$ is most probably
violated. Later, this statement can be made a little more
quantitative through
the investigation of the average transverse momentum square of quarks
and antiquarks in dependence of the longitudinal momentum fraction
$x$. But, before doing it, here demonstrate
realistic nature of the predictions of the CQSM, through the
analysis of the empirically well-known integrated quark and antiquark
distributions, the information of which is hidden in the
TMD distribution $f^{u+d} (x,\bm{k}_\perp)$.

\begin{figure}[!htb] \centering
\begin{center}
 \includegraphics[width=15.0cm]{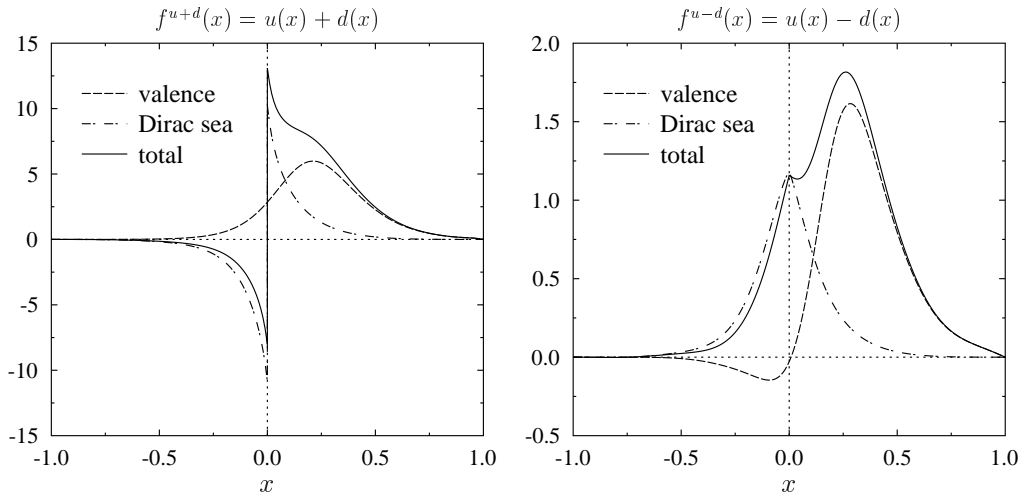}
\end{center}
\vspace*{-1.0cm}
\caption{The integrated unpolarized quark and antiquark distribution
function $f^{u+d}(x)$ with isoscalar combination obtained from
the corresponding TMD distribution via (\ref{Eq:integ}) (left panel).
The meaning of the curves is the same as in Fig.1.
Shown in the right panel is the unpolarized quark and
antiquark distribution with isovector combination, which has been
evaluated directly, i.e. without reference to the TMD distribution,
in the previous papers.}
\label{Fig:twist2pdf}
\end{figure}%

To demonstrate the reliability of the CQSM predictions for the unpolarized
TMD quark distributions shown in Fig.1 and Fig.2, we evaluate from
$f^{u+d} (x,\bm{k}_\perp)$ the corresponding
integrated quark distributions through the relation,
\begin{equation}
 f^{u+d} (x) \ = \ \int \,d^2 \bm{k}_\perp \,f^{u+d} (x,\bm{k}_\perp).
 \label{Eq:integ}
\end{equation}
The answer is shown in the left panel of Fig.\ref{Fig:twist2pdf},
which nicely reproduces the previous results for $u(x) + d(x)$
calculated directly without referring to the TMD quark
distribution \cite{Wakamatsu2003}. 
(A finer look however reveals that our new result for $f^{u+d}(x)$
shows a slight enhancement in the magnitude
of the Dirac-sea contribution as compared with our old result shown
in Fig.1(a) of \cite{Wakamatsu2003}, for which the check of the model space
dependence, i.e. the $D \rightarrow \infty$ and $k_{max} \rightarrow \infty$ limit, was not satisfactory enough.
The new result for the $\bar{d}(x) / \bar{u}(x)$ ratio
to be shown below is a little sensitive to this change of the isoscalar
unpolarized parton distribution.)
One confirms again that the proper introduction of the Dirac-sea
contribution is essential for reproducing the {\it positivity condition}
of the unpolarized antiquark distribution. One also sees that,
in the lower $x$ region, even the quark distribution is dominated
by the contributions of Dirac-sea quarks. If one compares this
figure of $f^{u+d}(x)$ with the Fig.\ref{Fig:ktf_pos}
for TMD distribution
$f^{u+d}(x,\bm{k}_\perp)$ especially at $x = 1.0 \times 10^{-6}$ and
$x = 0.1$, one can confirm that the longer range tail in $k_\perp$
of the Dirac-sea contribution is an important factor leading to its
strong enhancement in the small $x$ region, which we observe for the
integrated distribution $f^{u+d} (x)$.
In the right panel of Fig.\ref{Fig:twist2pdf}, we also show the
previously calculated isovector unpolarized quark distribution
$f^{u-d} (x)$, since it is necessary for the following 
comparison of the $\bar{d} (x) / \bar{u} (x)$ ratio.

\begin{figure}[htb] \centering
\begin{center}
 \includegraphics[width=15.0cm]{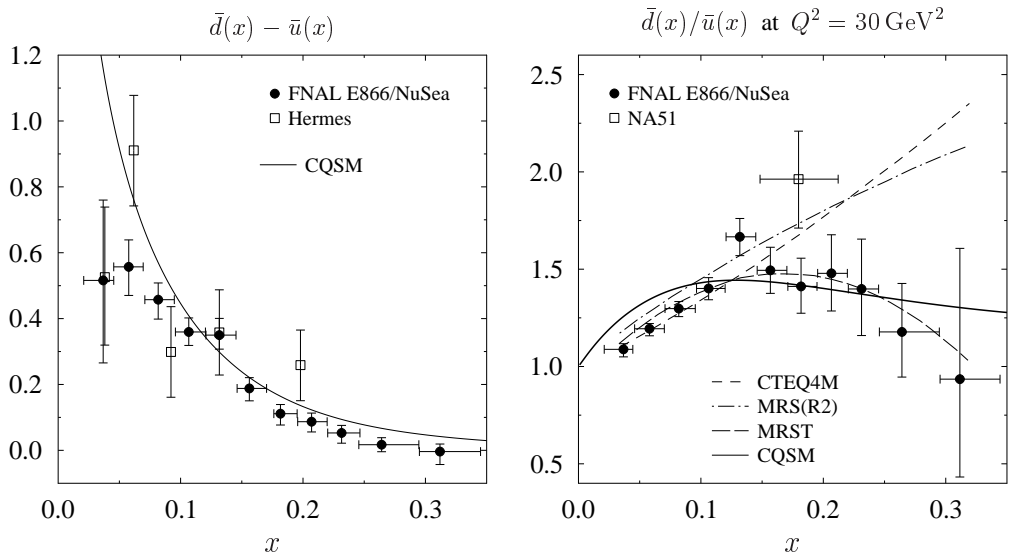}
\end{center}
\vspace*{-1.0cm}
\caption{The CQSM predictions for the difference function
$\bar{d}(x) - \bar{u}(x)$ in the proton in comparison with the
FNAL \cite{E866} and HERMES data \cite{HERMES1998} (left panel),
and for the ratio $\bar{d}(x) / \bar{u}(x)$
in comparison with the FNAL (Fermi National Accelerator Laboratory)
\cite{E866} and NA51 data \cite{NA51}
(right panel). The curves labeled CTEQ4M, MRS(R2) and MRST are
phenomenological PDF fits to the data before and after the E866
measurement.}
\label{Fig:asym_ubdb}
\end{figure}%

In order to demonstrate realistic nature of the CQSM predictions
for the antiquark distributions, we show
in Fig.\ref{Fig:asym_ubdb} its predictions for
the difference $\bar{u}(x) - \bar{d}(x)$ and the ratio
$\bar{d}(x) / \bar{u}(x)$ in comparison with the corresponding
empirical data.
As in the previous works \cite{WK1999},\cite{Wakamatsu2003},
the theoretical predictions for the
distribution functions at the high energy scales are obtained
by solving the DGLAP equations at the next-to-leading oder,
with the predictions of the CQSM as initial scale distributions
given at the scale $Q_{ini}^2 = 0.30 \,\mbox{GeV}^2$. 
(Here, the suffix ``ini'' should be understood as an
abbreviation of the word ``initial''.) 
A good agreement observed in the left panel
means that the model reproduces the famous NMC
observation,
i.e. the dominance of the $\bar{d}$-sea over the $\bar{u}$-sea
inside the proton, without any artificial fine-tuning. 
Shown in in the right panel of Fig.\ref{Fig:asym_ubdb} is the CQSM
prediction for the ratio
$\bar{d}(x) / \bar{u}(x)$ at $Q^2 = 30 \,\mbox{GeV}^2$
in comparison with the old NA51 data \cite{NA51} and the newer E866 data
extracted from the neutrino scatterings. 
The new MRST parton distribution function (PDF) fit including
the E866 data \cite{E866} together with the older CTEQ4M and MRS(R2) PDF
fits are also shown for reference.
One sees that the prediction of the CQSM is
qualitatively consistent with the E866 data as well as the new MRST
fit. We emphasize again that it is a completely parameter free
prediction of the model.

\begin{figure}[htb] \centering
\begin{center}
 \includegraphics[width=10.0cm]{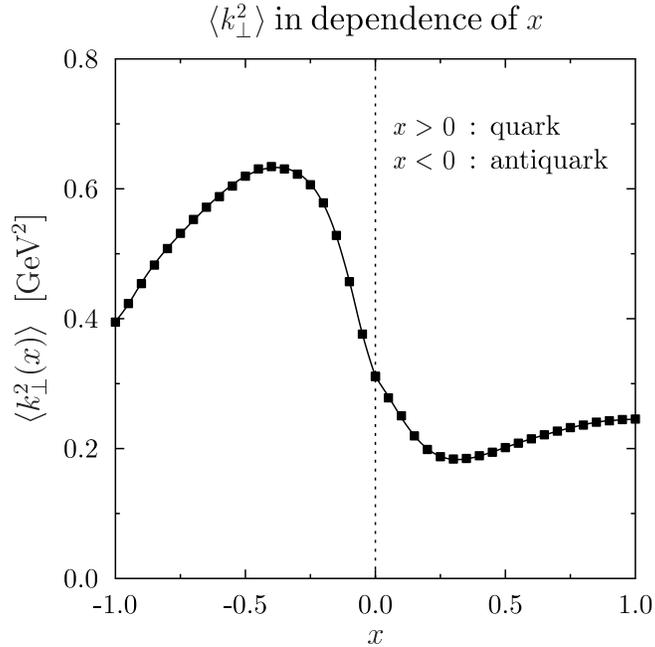}
\end{center}
\vspace*{-1.0cm}
\caption{The CQSM prediction for the average transverse momentum
square of quarks ($x > 0$) and antiquarks ($x < 0$) as a
function of the longitudinal momentum fraction $x$.} 
\label{Fig:abkt2}
\end{figure}

Now, after demonstrating that the CQSM well describes the fundamental
physics of unpolarized sea quark distributions, we come back to
its predictions on the unpolarized TMD distributions for quarks
and antiquarks.
From the already-given unpolarized TMD distributions
$f^{u+d} (x,\bm{k}_\perp)$ as functions of $x$ and $k_\perp^2$,
it is straightforward to evaluate the average transverse momentum
of quarks and antiquarks as a function of $x$ : 
\begin{equation}
 \langle k_\perp^2 (x) \rangle \ = \ 
 \frac{\int \,d^2 \bm{k}_\perp \, k_\perp^2 \,
 f^{u+d}(x,\bm{k})}{\int \,d^2 \bm{k}_\perp \,
 f^{u+d}(x,\bm{k})} .
\end{equation}
The resultant $\langle k_\perp^2 (x) \rangle$ is shown by
filled circles in Fig.\ref{Fig:abkt2}. The solid curve here
is a smooth fit to the numerical results by an 8th-order polynomial
as
\begin{equation}
 \langle k_\perp^2 (x) \rangle \ = \ \sum_{n = 0}^8 \,
 c_n \,x^n
\end{equation}
where
\begin{eqnarray}
 c_0 &=& 0.311122, \ \ c_1 \ = \ - 0.536064, \ \ 
 c_2 \ = \ - 2.4806, \nonumber \\
 c_3 &=& 18.7331, \ \ c_4 \ = \ - 41.7892, \ \ 
 c_5 \ = \ 40.5805, \nonumber \\
 c_6 &=& -11.3483, \ \ c_7 \ = \ - 7.48546, \ \ 
 c_8 \ = \ 4.26044, 
\end{eqnarray}
for $x > 0$, while
\begin{eqnarray}
 c_0 &=& 0.311122, \ \ c_1 \ = \ - 0.805606, \ \ 
 c_2 \ = \ 14.9032, \nonumber \\
 c_3 &=& 114.34, \ \ c_4 \ = \ 376.97, \ \ 
 c_5 \ = \ 693.994, \nonumber \\
 c_6 &=& 733.782, \ \ c_7 \ = \ 415.798, \ \ 
 c_8 \ = \ 97.7558, 
\end{eqnarray}
for $x < 0$.

One clearly sees that the average transverse momentum square is
strongly dependent on the longitudinal momentum fraction $x$
of quarks and antiquarks.
This reconforms that the frequently used assumption of factorization
in the variables $x$ and $k_\perp$ is significantly broken.
(The plausible breakdown of the usual Gaussian ansatz for a
factorized $k_\perp$ dependence was also indicated in the
model calculation by Pasquinni et al. \cite{PCB2008})
Very recently, the first lattice QCD study of the transverse
momentum distributions of quarks in the nucleon has been
reported \cite{LHPC2008}.
The validity of factorization in $x$ and $k_\perp$ was examined,
and it has been concluded that the factorization hypothesis holds
within the statistics of the simulation in sharp contrast to our
present analysis based on the CQSM.
In our opinion, the results of the lattice QCD simulation at the
present level cannot be taken too seriously by
several reasons. First, the simulation is preliminary in that it
was performed in the fictitious heavy-pion region around
$m_\pi \simeq 500 \, \mbox{MeV}$. What is lacking here is the
nonperturbative chiral dynamics of light quarks, which is the
core of the celebrated NMC observation, or the physics of
Gottfried sum.  
Second, the test of factorization is far from persuasive,
since the evaluation of full $x$-dependence
is beyond the scope of the presently-known algorithms of
lattice QCD.

Turning back to the $x$-dependence of the average transverse momentum
square of quarks and antiquarks, we continue to analyze the
prediction of the CQSM. Let us first look into
the positive $x$ region corresponding to the quark distribution.
Very strangely, the average transverse momentum square has a minimum
around $x \sim 0.25$, and it increases as $x$ becomes larger.
This feature can be understood as follows.
Around $x \sim (0.2 - 0.3)$, the integrated quark distribution
$f^{u+d}(x)$ has a peak, which is dominated by the contribution of the
valence-level quarks. This means that the relative importance of
Dirac-sea quarks, which has higher $k_\perp$ components, as compared
to the valence-level quark, is smallest around there, which explains
the above-mentioned feature of $\langle k_\perp^2 (x) \rangle$ as
a function of $x$.
One may also notice that the average transverse momentum square around
$x = 0$ is fairly large, which indicates that the quarks (and antiquarks)
with small longitudinal momentum fraction would carry sizable amount
of orbital angular momentum along the $z$-axis, i.e. along the
the direction of nucleon spin.
We recall that this observation is consistent with the results
of \cite{WW2000}, in which a direct calculation of quark
orbital-angular-momentum distributions was carried out.
(See also the analyses of the quark orbital angular momentum
from the viewpoint of the
generalized parton distribution functions \cite{OPSUG2005}
\nocite{WT2005}\nocite{WY2006}-\cite{WY2008}.)

Another unique feature of the CQSM prediction is that the
magnitude $\langle k_\perp^2 (x) \rangle$ is much larger
in the negative $x$ region, corresponding to the antiquarks.
Another way of convincing this feature is to compare the two
quantities defined below :
\begin{eqnarray}
 \langle k_\perp^2 \rangle^Q &\equiv& 
 \frac{\int_0^1 \,dx \,\langle k_\perp^2 (x) \rangle \,f^{u+d} (x)}
 {\int_0^1 \,dx \,f^{u+d} (x)}, \\
 \langle k_\perp^2 \rangle^{\bar{Q}} &\equiv& 
 \frac{\int_{-1}^0 \,dx \,\langle k_\perp^2 (x) \rangle \,f^{u+d} (x)}
 {\int_{-1}^0 \,dx \,f^{u+d} (x)} \ = \ 
 \frac{\int_0^1 \,dx \,\langle k_\perp^2 (x) \rangle \,
 f^{\bar{u}+\bar{d}} (x)}
 {\int_0^1 \,dx \,f^{\bar{u}+\bar{d}} (x)},
\end{eqnarray}
which represent the average transverse momentum square
for quarks and antiquarks, respectively. Numerically, we find that
\begin{eqnarray}
 \langle k_\perp^2 \rangle^Q &=& 0.224 \,\mbox{GeV}^2, \\
 \langle k_\perp^2 \rangle^{\bar{Q}} &=& 0.445 \,\mbox{GeV}^2,
\end{eqnarray}
which clearly shows that the average transverse momentum of
antiquarks is much larger than that of quarks.
We can also estimate the average transverse
momentum square of quarks and antiquarks altogether from
\begin{equation}
 \langle k_\perp^2 \rangle^{Q + \bar{Q}} \ \equiv \ 
 \frac{\int_{-1}^1 \,dx \,\langle k_\perp^2 (x) \rangle \,f^{u+d} (x)}
 {\int_{-1}^1 \,dx \,f^{u+d} (x)},
\end{equation}
which gives
\begin{equation}
 \langle k_\perp^2 \rangle^{Q + \bar{Q}} \ = \ 0.266 \,\mbox{GeV}^2.
\end{equation}
This value of average transverse momentum square is remarkably close
to that used in the phenomenological analysis of the semi-inclusive
reactions \cite{ABCDEGKMMMPSVY2005}.
Note, however, that the average transverse momentum
is a scale dependent quantity, which is believed to grow as
$Q^2$ increases.
The predictions of the CQSM corresponds to the scale
$Q^2 \sim (0.30 - 0.40) \,\mbox{GeV}^2$
\cite{WK1999},\cite{Wakamatsu2003}, while the phenomenological
analysis in \cite{ABCDEGKMMMPSVY2005} 
should correspond to somewhat higher energy scale.
In any phenomenological analysis, one must keep in mind this
scale dependent nature of the average transverse momentum square
of quarks and antiquarks.

Before ending this section, we make a short comment on the
so-called transverse-coordinate representation of the
TMD parton distributions.
The $Q^2$-evolution of the TMD parton distributions or the
unintegrated parton distributions is described by the
Catani-Ciafaloni-Fiorani-Marchesini (CCFM)
equations \cite{Ciafaloni1988}\nocite{CFM1990}-\cite{BCM1980}.
A nice feature of the CCFM equation at the leading order 
is that they reproduce the convectional leading-order
Dokshitzer-Gribov-Lipatov-Altarelli-Parisi (DGLAP) equations for
the integrated distributions. The CCFM equations are known to take
a particularly  simple structure if one introduces the
so-called transverse-coordinate representation of the parton
distributions \cite{KT1982}\nocite{DDT1978}
\nocite{Kwiecinski2002}-\cite{GK2003}.
To explain it, we represent below the
TMD quark (or antiquark) distribution of flavor $a$ as
$f^a (x, k_\perp, Q^2)$ with $k_\perp = | \bm{k}_\perp |$,
where the dependence on the scale $Q^2$ is also
shown explicitly. The transverse-coordinate representation
$\bar{f}^a (x, b, Q^2)$ of a quark (or an antiquark) of flavor
$q$ is introduced as the two-dimensional Fourier transform of
the corresponding TMD distribution $f^a (x, k_\perp, Q^2)$ : 
\begin{equation}
 \bar{f}^a (x, b, Q^2) \ \equiv \ \int \,d^2 \,\bm{k}_\perp \,\,
 e^{\,i \,\bm{b} \cdot \bm{k}_\perp} \,\,f^a (x, k_\perp, Q^2),
\end{equation}
where $\bm{b}$ is the transverse coordinate vector. Clearly,
at $b = 0$, the function $\bar{f}^a (x, b, Q^2)$ reduces to the familiar
integrated distribution $f^a (x, Q^2)$. The greatest advantage of using
the transverse-coordinate representation is that the corresponding
evolution equations for the unintegrated distributions,
$\bar{f}_{NS} (x, b, Q^2), \bar{f}_{S} (x, b, Q^2)$ and
$\bar{f}_g (x, b, Q^2)$, become diagonal in the transverse
coordinate $b$. Here, just for illustration, we show the
evolution equation for the nonsinglet (NS) distribution.
It takes the form \cite{GKB2003},\cite{AB2004} :
\begin{eqnarray}
 Q^2 \,\frac{\partial^2}{\partial Q^2} \,\,
 \bar{f}_{NS} (x, b, Q^2) &=& \frac{\alpha_S (Q^2)}{2 \,\pi} \,
 \int_0^1 \,dz \,\,P_{qq} (z) \\
 \times \!\!\!\!\! &\,& \!\!\!\!\!
 \left\{\,\Theta (z-x) \,J_0 [ (1-z) \,Q \,b] \,
 \bar{f}_{NS} \left( \frac{x}{z}, b, Q^2 \right) \ - \ 
 \bar{f}_{NS} (x, b, Q^2) \,\right\} ,
\end{eqnarray}
which certainly is diagonal in the variable $b$.
Once these evolution equations are solved in a similar way as the
standard DGLAP equations, the average transverse momentum square
for a given distribution at any $Q^2$ can be evaluated from
\begin{equation}
 \langle k_\perp^2 (x, Q^2) \rangle^a \ = \  
 \frac{\int \,d^2 \bm{k}_\perp \,\,k_\perp^2 \,\,f^a (x, k_\perp, Q^2)}
 {\int d^2 \bm{k}_\perp \,\,f^a (x, k_\perp, Q^2)} \ = \ 
 - \,4 \,\left.
 \frac{\frac{d}{d b^2} \,\bar{f}^a (x, b, Q^2)}
 {\bar{f}^a (x, b, Q^2)} \,\right|_{b=0} .
\end{equation}
The transverse-coordinate representation of the unintegrated parton
distribution discussed above should not be confused with more
popular impact-parameter dependent parton distributions introduced
by Burkardt \cite{Burkardt2000}, although there should be strong
correlation between them.
The simplest impact-parameter dependent parton distribution
function $f^a (x, b_\perp)$ is the two dimensional Fourier
transform of the unpolarized generalized parton distribution
function $H^a (x, \xi, t)$ with zero skewdness parameter $\xi = 0$ :
\begin{equation}
 f^a (x, b_\perp) \ = \ \int \,\frac{d^2 \bm{\Delta}}{(2 \,\pi)^2} \,\,
 e^{\,- \, i \,\bm{\Delta} \cdot \bm{b}_\perp} \,\,
 H^a (x, \xi = 0, t = - \,\bm{\Delta}^2) .
\end{equation}
We recall that the frequently used parametrization \cite{DFJP2005} of
$H^a (x, \xi=0, t)$ is of non-factorizable form in $x$ and $t$ as
\begin{equation}
 H^a (x, \xi = 0, t) \ = \ f^a (x) \,\exp \,\left[\,
 \beta^a (x) \,t \,\right] ,
\end{equation}
with
\begin{equation}
 \beta^a (x) \ = \ \left[ \,\alpha^\prime \,\ln (1/x) + B_a \,\right] \,
 (1 - x)^3 \ + \ A_a \,x \,(1 - x)^2 ,
\end{equation}
which means that $f^a (x, b_\perp)$ is also non-factorizable in
$x$ and $b_\perp$.
This in turn strongly indicates that the transverse-coordinate
representation of the unintegrated parton distributions, and
consequently, the TMD parton distributions are most likely to be
non-factorizable in the variables $x$ and $k_\perp$, as our
predictions based on the CQSM shows explicitly.

\section{Conclusion \label{Sect:conclusion}}

To conclude, we have reported the first calculation of the simplest
but most fundamental TMD parton distribution in the nucleon,
i.e. the unpolarized TMD quark and antiquark distributions with
isoscalar combination, within the framework of the CQSM.
The realistic nature of the CQSM predictions
was demonstrated by showing that the integrated unpolarized
distribution $f^{u+d} (x)$, obtained from the corresponding TMD
distribution $f^{u+d} (x,\bm{k}_\perp)$, is consistent with
the available empirical information on the light-flavor unpolarized
parton distributions. It was found that the predicted average
transverse momentum square of quarks and antiquarks depends
strongly on their longitudinal momentum fraction $x$.
We also estimate the average transverse momentum of quarks and
antiquarks separately to find, somewhat unexpectedly, that
$\langle k_\perp^2 \rangle^Q \simeq 0.224 \,\mbox{GeV}^2$ and
$\langle k_\perp^2 \rangle^{\bar{Q}} \simeq 0.445 \,\mbox{GeV}^2$,
that is, the antiquarks have much higher average transverse momentum
than the quarks. On the other hand, the average momentum transfer of
quarks and antiquarks altogether turns out to be
$\langle k_\perp^2 \rangle^{Q + \bar{Q}} \simeq 0.266 \,\mbox{GeV}^2$,
which is order of magnitude consistent with the recent
phenomenological analysis, although we must be careful about the fact
that the average transverse momentum square is a $Q^2$-dependent
quantity.




\end{document}